\documentclass[aps,pre,preprint,double-spaced,floatfix,showpacs]{revtex4}

\usepackage[dvips]{graphicx}
\usepackage{subfigure}
\usepackage{epsfig}
\usepackage[english]{babel}
\usepackage{amsmath}
\usepackage{amssymb}
\usepackage{times}

\begin{document}

\author{Jin W. Wang and Thomas A. Witten}

\affiliation{The James Franck Institute and The Department of Physics,
The University of Chicago, 929 East 57th Street, Chicago, Illinois 60637}

\title{The compensation of Gaussian curvature in developable cones is local}

\date{\today}

\begin{abstract}
In this paper we use the angular deficit scheme [V. Borrelli, F. Cazals, and J.-M. Morvan, {\sl Computer Aided Geometric Design} {\bf 20}, 319 (2003)] to determine the distribution of Gaussian curvature in developable cones (d-cones) [E. Cerda, S. Chaieb, F. Melo, and L. Mahadevan, {\sl Nature} {\bf 401}, 46 (1999)] numerically. These d-cones are formed by pushing a thin elastic sheet into a circular container. Negative Gaussian curvatures are identified at the rim where the sheet touches the container. Around the rim there are two narrow bands with positive Gaussian curvatures. The integral of the (negative) Gaussian curvature near the rim is almost completely compensated by that of the two adjacent bands. This suggests that the Gauss-Bonnet theorem which constrains the integral of Gaussian curvature globally does not explain the spontaneous curvature cancellation phenomenon [T. Liang and T. A. Witten, {\sl Phys. Rev. E} {\bf 73}, 046604 (2006)]. The locality of the compensation seems to increase for decreasing d-cone thickness. The angular deficit scheme also provides a new way to confirm the curvature cancellation phenomenon.

\end{abstract}


\maketitle
\section{INTRODUCTION}
\begin{figure}[h!]
  \begin{center}
  \includegraphics[width=0.8\textwidth]{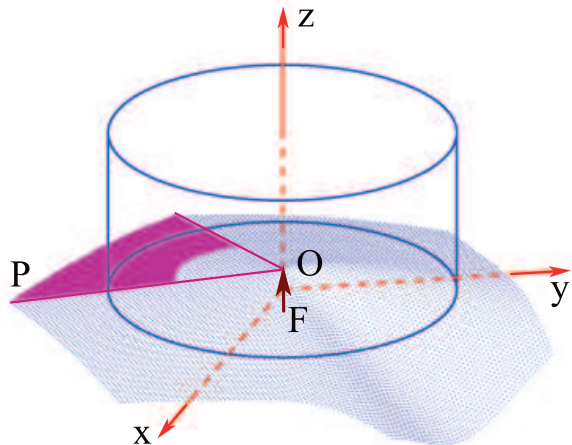}
  \caption{A typical simulated d-cone formed by pushing the center of a hexagonal elastic sheet $ O $ against a cylindrical container with force $ F $. It has side length $l=60a$, container radius $R=38a$, deflection $ \epsilon=0.095$ and thickness $ h=0.102a $, where $ a $ is the lattice spacing. $ OP $ is one boundary of the 60 degree sector opposite to the buckled region. Gaussian curvatures are integrated over the shaded region in Sec. III. The cylinder is represented by a circular potential shown in Sec. II.}
  \label{dcone}
  \end{center}
\end{figure}

Over the past few years, there has been a marked and still increasing interest in the geometrical and mechanical properties of developable cones or d-cones, illustrated in Figure~\ref{dcone}. The d-cone offers an especially simple example of focusing of mechanical  stress in a thin elastic sheet owing to the interplay of bending and stretching energy deformation.  Several puzzles regarding the scaling of this focusing remain unresolved\cite{witten1}. Chief among these is the observed vanishing of mean curvature at the supporting rim of the d-cone. This paper investigates the possible role of the Gauss-Bonnet theorem in explaining this vanishing phenomenon.

\begin{figure}[h!]
  \begin{center}
  \includegraphics[width=0.5\textwidth]{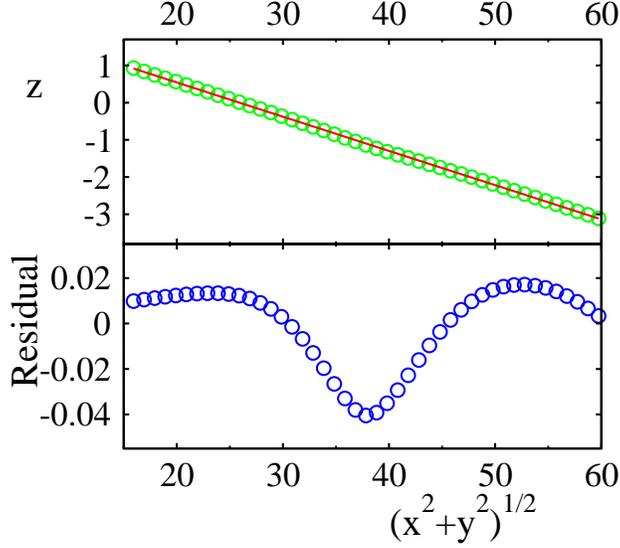}
  \caption{Top: The radial profile of the d-cone along $ OP $. The slope of the fitted line is -0.092. Bottom: The residual plot of difference between the data points and the fitted line. This plot clearly shows that radial curvatures are not zero, especially near the rim where $ \sqrt{x^2+y^2}=38 $.}
  \label{kink}
  \end{center}
\end{figure}

In principle the shape of a d-cone, including the focusing phenomena noted above, may be found by solving the von K\'{a}rm\'{a}n equations describing large deformations of thin plates\cite{amar1}. Cerda and Mahadevan\cite{cerda1}  analyzed the geometry of a single d-cone formed by pushing a thin elastic circular sheet into a cylindrical frame by applying a centered transverse force directed along the axis of the cylinder, in the limit of pure bending. They obtained the shape of the d-cone by minimizing the bending energy and showed that the aperture angle of the buckled part has a universal value of 139 degrees and that the sheet exerts a concentrated point force to the frame at the two take-off points where the sheet bends away from the frame. This finding was confirmed by Liang and Witten\cite{liang1}. 

These authors\cite{liang2} also uncovered a striking spontaneous curvature cancellation phenomenon as noted above. Though the ideal shape of a d-cone has zero radial curvature $ C_{rr} $ where $ r $ and $ \theta $ are defined as radial and angular components in the material coordinate system, the real shape must have nonzero radial curvature at the rim, demonstrated in Figure \ref{kink}. The localized force requires an outward radial curvature of the surface so that the surface elements at the frame may maintain equilibrium. As the thickness of the elastic sheet goes to zero, they found that within the numerical accuracy, the radial curvature $C_{rr}$ and the azimuthal curvature $C_{\theta \theta}$ are equal and opposite at the rim, so that the mean curvature virtually vanishes there. Though it arises from mechanical equilibrium, the phenomenon is purely geometric: it does not involve material parameters\cite{liang2,witten1}. The von K\'{a}rm\'{a}n equations must be able to account for the cancellation effect, but this has not been done yet. 

Any outward radial curvature $C_{rr}$ combined with the inward azimuthal curvature $C_{\theta \theta}$ necessarily creates a nonzero Gaussian curvature $K = C_{rr} C_{\theta \theta}$. The vanishing of mean curvature $(C_{rr} + C_{\theta \theta})/2$ means that $K $ at the rim approaches a nonzero constant as the thickness of the sheet goes to zero.  This  contrasts with the overall behavior of thin sheets, which approach the isometric, developable shape for which it must go to zero. 

The vanishing mean curvature at the rim amounts to a nonlocal geometrical constraint. It is nonlocal because it depends on the overall geometry of the sheet.  The curvature does not vanish when one modifies this geometry by cutting the sheet radially or by replacing the flat sheet by a cone\cite{liang2}. The effect also depends on the nonlocal forcing at the vertex. Given the importance of nonlocal geometry, it is natural to look for a connection with the main known geometric constraint on curvature: namely the Gauss-Bonnet theorem. This theorem states that\cite{struik1} the integral of Gaussian curvature over a region $ M $ of a surface is related to the integral of the geodesic curvature $ \kappa_g $ over the boundary of that region through $ \int_{M}K\, dA +\int_{\partial M}\kappa_{g}\, ds =2\pi $.  The geodesic curvature for a point on a curve lying on a surface is defined as the curvature of the orthogonal projection of the curve onto the tangent plane to the surface at the point. As noted below, there is an exact counterpart of the theorem for polyhedral surfaces.

The Gauss-Bonnet theorem implies that any region of net negative Gaussian curvature, such as the d-cone rim, must be compensated by a region of positive curvature elsewhere in the surface or a change in the boundary integral. Our aim in this paper is to determine where the rim curvature is actually compensated. Since the vanishing mean curvature is  the result of global forces in the sheet, it would be natural to find that the rim Gaussian curvature is compensated by positive Gaussian curvature concentrated in the core region. Such nonlocal compensation arises, for example, if one pushes a stretched membrane such as a drum head at an interior point. Here positive Gaussian curvature appears at the forcing point. Under weak deformation, as one moves away from the forcing point, the deformation becomes conical, with no Gaussian curvature. The compensating negative Gaussian curvature appears only at the constrained boundary of the membrane--the same place where the point force is compensated. It is natural to expect analogous behavior for the d-cone.

\begin{figure}
\centering
\begin{tabular}{cc}
\epsfig{file=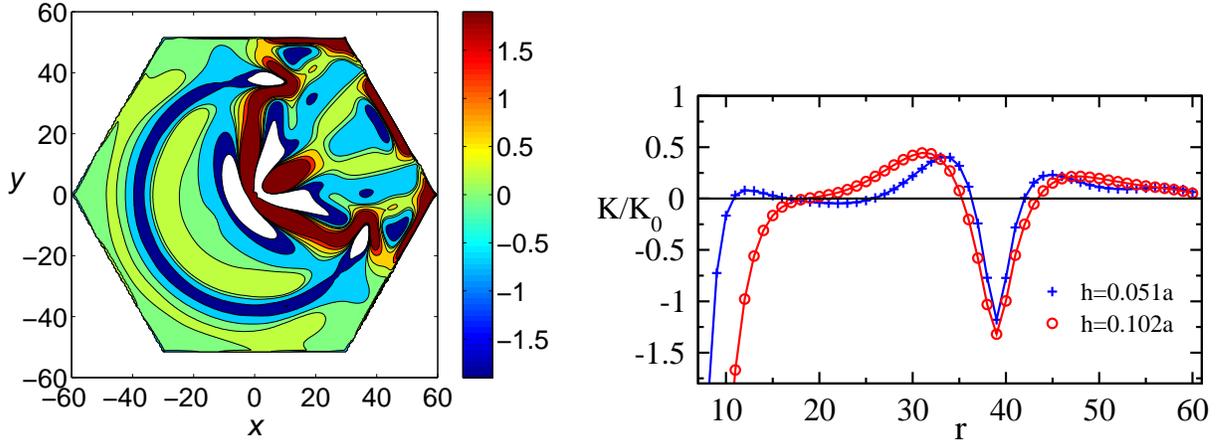,width=0.5\linewidth ,clip=} &
\epsfig{file=3b.eps,width=0.5\linewidth ,clip=} 
\end{tabular}
\caption{Left: Density plot of the Gaussian curvature $ K $ in the d-cone shown in Figure \ref{dcone}. The point $P$ is at the lower left corner. The gaussian curvature at each point is indicated by the scale at right.  White areas indicate off-scale positive or negative values of $K$. Right: The radial profiles of $ K $ as a function of $ r $ for two different values of thickness $ h $. $ K $ is normalized by $ K_0 $ which is defined in Sec. II.}
\label{radial1}
\end{figure}

These Gauss-Bonnet integrals are expected to be influenced by the elastic thickness $h$ of the sheet. As this thickness goes to zero relative to the size $R$ of the container, the sheet approaches the isometric state in which the energy penalty for stretching becomes arbitrarily large. In this limit the Gaussian curvature must approach zero at almost every point, since Gaussian curvature entails stretching\cite{millman1}. Thus the negative contribution to the Gauss-Bonnet integral near the rim must go to zero. Although this negative contribution becomes small, it must nevertheless be compensated somewhere on the surface. The compensation could occur far from the rim or close to it. We may investigate this question by evaluating the integral of $ K $ in a band near the rim. If the nagative region near the rim is nearly compensated by a positive region within the band, then $\int_{band} K \ll \int_{band}|K|$. If instead the compensation occurs largely outside the band, then $ \int_{band}K / \int_{band}|K| $ remains finite as the thickness goes to zero.

Our numerical study reported below finds that the rim Gaussian curvature is compensated locally, not globally. There is a narrow annulus near the rim where the Gauss-Bonnet integral nearly vanishes, due to nearly cancelling positive and negative contributions. There is an analogous local region around the core. The locality appears to increase for decreasing membrane thickness. Our methods allow us to confirm the vanishing-mean-curvature phenomenon in a new way, using the Gaussian curvatures. The numerical models we used are presented in Sec. II. In Sec. III, we describe our numerical findings in detail. In Sec. IV we discuss limitations and implications of our findings.

\section{NUMERICAL METHODS}
\subsection{Numerical Model}

We model an elastic sheet by a triangular lattice of springs of un-stretched length $a$ and spring constant $k$ after Seung and Nelson\cite{seung1}. Bending rigidity is introduced by assigning an energy of $J(1-\hat{n}_{1} \cdot \hat{n}_{2})$ to every pair of adjacent triangles with normals $ \hat{n}_{1} $  and $ \hat{n}_{2} $. When strains are small compared to unity and radii of curvature are large compared to the lattice spacing $ a $, this model is equivalent to an elastic sheet of thickness $ h=a \sqrt{8J/k} $ made of an isotropic, homogeneous material with bending modulus $ \kappa = J\sqrt{3}/2 $, Young's modulus $ Y=2ka/h\sqrt{3} $ and Poisson's ratio $ \nu =1/3 $. The lattice spacing $ a $ is set to be 1. The shape of the sheet in our simulation is a regular hexagon of side length $ R_p $. The typical value of $ R_p $ is $ 60a $, as indicated in Figure \ref{dcone}. 

To obtain a single d-cone shape, we need to simulate the constraining container rim and pushing force. The rim lies in the $ x-y $ plane and is described by equation $ x^2+y^2=R^2 $, where $ R $ is the radius of the container. The deflection of the sheet is defined as $ \epsilon\equiv d/R $, where $ d $ is the distance by which the center of the sheet has been pushed into the container. A d-cone with deflection $ \epsilon $ touches the container at distance $ R \sqrt{1+\epsilon ^2} $ away from the apex. For each d-cone, we define a constant $ K_0 $ as $ K_0=(\epsilon /(R \sqrt{1+\epsilon ^2}))^2 $, where $ \epsilon /(R \sqrt{1+\epsilon ^2}) $ is the azimuthal curvature $C_{\theta \theta}$ of a cone with opening angle $ 2\tan^{-1}(d/R)$ at $ r= R \sqrt{1+\epsilon ^2}$. In a real d-cone, $ K_0 $ is slightly different from the square of the azimuthal curvature at the rim due to lattice effect and the finite thickness $ h $ of the sheet. $ K_0 $ should approach $ C_{\theta \theta}^2 $ at the rim as $ h $ goes to zero. Pushing in the center of the sheet is accomplished by introducing a repulsive potential of the form $ U_{\text{force}}(x_1, y_1, z_1)=- (z_1+a)G(x_1,y_1)F $, where $  G(x_1, y_1)=[(1+(x_1/\xi)^2)(1+(y_1/\xi)^2)]^{-1}$, $x_1 $,  $y_1 $ and $ z_1 $ are the coordinates of the lattice point in the center, $ \xi $ is a constant of order $ 0.1a $ and $ F $ is the magnitude of the pushing force. $ G(x_1, y_1) $ is introduced to make sure that when the sheet is being pushed the lattice point in the center does not stray away from the axis of the cylindrical container, i.e. $(x_1,y_1)\simeq (0,0)$. The constraining rim is implemented by a potential of the form $ U_{\text{rim}}=\sum C_{p}/[(\sqrt{x_i^2+y_i^2}-R)^2+z_i^2]^4 $, where $ C_p $ is a constant and the summation is over all lattice points with coordinates $ (x_i, y_i, z_i) $. The value of $ C_p $ is choosen so that the shortest distance between the lattice points and the rim is close to one lattice spacing. 

The conjugate gradient algorithm\cite{press1} is used to minimize the total elastic and potential energy of the system as a function of the coordinates of all lattice points. This lattice model behaves like a continuum material provided that the curvatures are everywhere sufficiently small compared to $ 1/a $. This restricts the values of deflection $ \epsilon $ of d-cone to be below 0.25\cite{liang2}. $ \epsilon $ is between 0.09 and 0.15 in our simulations.

\subsection{Evaluation of curvatures}

There are two ways to determine the curvatures. 

One way, denoted as Scheme 1, is to obtain the curvature tensor from each triangle in the sheet. Once the curvature tensor is known, Gaussian curvature and mean curvature are the determinant and average of the diagonal elements of the tensor, respectively. For this scheme to work, we have to take the curvature tensor to be constant across each triangle. We calculate it following \cite{didonna1} using the relative heights of the six vertices of the three triangles that share a side with the given triangle.

Gaussian curvatures can also be obtained by using the angular deficit scheme\cite{borrelli1} based on the Gauss-Bonnet Theorem. The angular deficit scheme approximates the Gaussian curvature of a smooth surface $S$ from the angular deficit of an inscribed polyhedral surface. The angular deficit of a vertex of a polyhedron is the amount by which the sum of the angles of the faces at the vertex falls short of $ 2\pi $. Let $ p $ be a vertex of a polyhedral surface and let its nearest neighbors to be $ p_i, i=1,\ldots,n $ in clockwise direction. In triangle $ p~p_{i}p_{i+1} $, let $ \gamma_{i} $ denote the angle at $ p $, i.e., $ \angle p_{i}pp_{i+1} $ , with $ p_{n+1}=p_1 $. The angular deficit at $ p $ is defined as $ 2\pi-\sum_{i=1}^{n} \gamma_{i} $. In our equilateral triangular lattice model, if a vertex $ p $ is not on the edge of the polyhedral surface, then $ n=6 $, and the angular deficit and the Gaussian curvature $ K $ of the smooth surface $ S $ at $ p $ satisfy:
\begin{equation}
K  = (2\pi - \sum_{i=1}^{6} \gamma_{i})/(\frac{\sqrt{3}}{2}a^2) = (\sum_{i=1}^{6} (\dfrac{\pi}{3} - \gamma_{i}))/(\frac{\sqrt{3}}{2}a^2) \label{angular_curvature}
\end{equation} 
in the limit of small lattice spacing\cite{borrelli1}. Since angular deficits can be easily calculated in our triangular lattice model, this is a convenient way to evaluate Gaussian curvatures. For a vertex $ p $ on the edge, $ n=3\text{ or }4 $. The angular deficit at $ p $ is $ \sum (\pi/3 -\gamma_i) $ where we restrict the angles to those subtending the sheet. After this adjustment, the sum of the angular deficits over all the vertices on the sheet is equivalent to the sum of angular deficits over all the triangles and should be exactly zero. In our simulation, the magnitude of this sum is on the order of $10^{-10}$, which is consistent with zero given our numerical round-off errors.

Equation (\ref{angular_curvature}) can be seen as the counterpart of the Gauss-Bonnet theorem on smooth surface for polyhedral surfaces. In the polyhedral case, the Gaussian curvature of the surface is concentrated on the vertices. It is zero on faces and edges. By approximating the polyhedron by a sequence of smooth surfaces, one verifies that $ K $ consists of $ \delta $-function contribution at each vertex $ j $ each being equal to its angular deficit: $ K(\vec{r})=\sum_j [\sum_i (\pi /3 -\gamma_{ij})]\delta ^2 (\vec{r}-\vec{r}_j) $. This special case of Gaussian-Bonnet theorem states that the sum of the angular deficits over all the vertices of a polyhedron is $ 2\pi \chi $, where $ \chi $ is the Euler characteristic of the polyhedron. For example, if a polyhedron is homeomorphic to a sphere, then its Euler characteristic $ \chi=2 $ and the total angular deficits of all its vertices is $4\pi $. So Equation (\ref{angular_curvature}) simply states that in our regular triangular lattice, the Gaussian curvature of a smooth surface can be approximated by the normalized Gaussian curvature of the discrete lattice. There is great temptation to generalize this to other kinds of lattices, but it is only true when the geometry of the lattices is precisely controlled\cite{borrelli1}.

These two schemes of obtaining the Gaussian curvature give consistent values everywhere on the d-cone except in the core region where the high curvatures invalidate the smoothness assumption of Scheme 1. All Gaussian curvatures presented in subsequent sections are those of our polyhedral surface determined by the angular deficit scheme, the robustness and accuracy of which has been tested extensively\cite{borrelli1,xu1,surazhsky1}. 

\section{RESULTS}

Numerically, Figure ~\ref{radial1}  shows the Gaussian curvature profiles along a radial line $ OP $ that is $ 5\pi/6 $ away from the symmetrical axis of the buckled region as indicated in Figure \ref{dcone}. Negative Gaussian curvatures are observed around $ r=38a $ in the material coordinate, where the sheet touches the confining rim, just as stated in the Introduction. This negative Gaussian curvature decays as one moves inward or outward from the rim, but the decay is not monotonic. It first goes to zero and then turns positive before approaching zero again. As expected, Gaussian curvatures change very dramatically in the core region. 

\begin{figure}
  \includegraphics[width=0.6\textwidth]{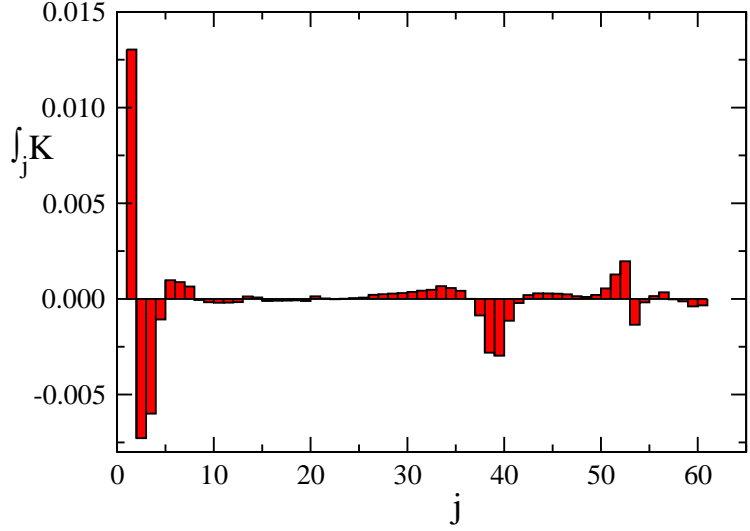}
  \caption{The integral of Gaussian curvature $ \int_{j} K $ within each band $ A_j $ as a function of $ j $, where $ j $ represents the material distance from band $ A_j $ to the center of the d-cone.}
  \label{whole}
\end{figure}

The existence of two regions with positive Gaussian curvatures around the contact region seems to indicate that the negative Gaussian curvatures in the contact region is compensated locally. To test this, on the d-cone with thickness $ h=0.102a $ we integrate the Gaussian curvature over a band within the unbuckled region shown as the shaded region in Figure \ref{dcone}. The results are as follows:
\begin{eqnarray}
\int_{\theta=5\pi/6}^{7\pi /6} \int_{r=20a}^{59a} K(\theta , r) r\,dr\,d\theta = -5.5\times 10^{-5}, \\
\int_{\theta=5\pi/6}^{7\pi /6} \int_{r=20a}^{59a} \vert K(\theta , r)\vert r\,dr\,d\theta = 2.8\times 10^{-3} \,.
\end{eqnarray}
These integrals of the Gaussian curvature are two orders of magnitude smaller than the integrals of the absolute value of the Gaussian curvature, which means that the negative Gaussian curvature at the rim is effectively compensated by the positive Gaussian curvature close to the rim, i.e., the Gaussian curvature is compensated locally. 

By comparing the radial profiles of the Gaussian curvature in Figure ~\ref{radial1}, one would notice that the negative Gaussian curvature at the rim decays to zero within a shorter distance on the thinner d-cone. Also, on the thinner d-cone the range of two regions of positive Gaussian curvature is correspondingly smaller. This suggests that the locality of the Gaussian curvature compensation increases as the thickness of the d-cone decreases.

The local compensation of Gaussian curvature applies not only to the unbuckled region, but also to the whole d-cone. We divide a d-cone into circular bands with each band $ A_j $ containing lattice points that satisfy $ (j-1)\leq r < j, \;  j=1, \ldots, 61$, where $ r $ is the material distance from the lattice point to the center of the d-cone. Then within each band $ j $, we integrate the Gaussian curvature and denote the integral as $ \int_{j}K $. This $ \int_{j}K $ is plotted against $ j $ in Figure ~\ref{whole}. From Figure~\ref{whole} We can roughly decompose the d-cone to 3 major regions, the core region with $ 0 \leq r \leq 20a $, the rim region with $ 20a < r \leq 50a $ and the edge region with $ 50a < r \leq 60a$ where the integrals are primarily contributed by the buckled region. 

\begin{figure}
  \includegraphics[width=0.6\textwidth]{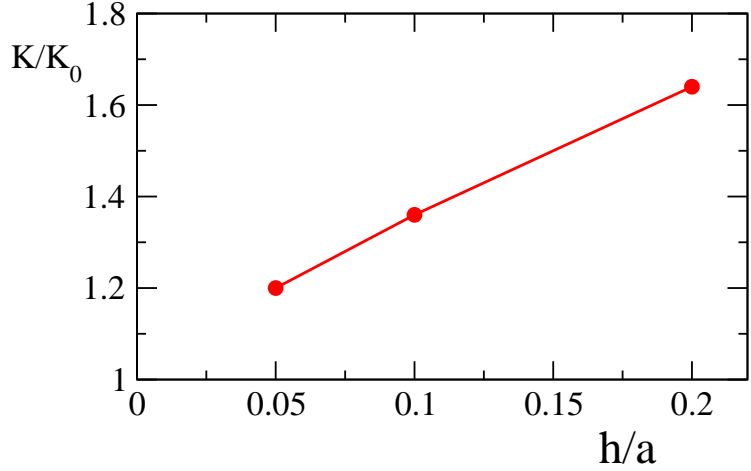}
  \caption{The normalized average gaussian curvature $ K $ at the rim as a function of the thickness of the sheet $ h $. Gaussian curvatures at rim points were inferred from interpolation. These interpolated $K$ values were then averaged over the contacting rim.}
  \label{Kvsh}
\end{figure}

In the core region, we have
\begin{eqnarray}
\int_{\theta=0}^{2\pi} \int_{r=0}^{20a} K(\theta , r) r\,dr\,d\theta = 2.5\times 10^{-4}, \\
\int_{\theta=0}^{2\pi} \int_{r=0}^{20a} \vert K(\theta , r)\vert r\,dr\,d\theta = 1.0\times 10^{-1} \,,
\end{eqnarray}
and in the rim region,
\begin{eqnarray}
\int_{\theta=0}^{2\pi} \int_{r=20a}^{50a} K(\theta , r) r\,dr\,d\theta = -1.5\times 10^{-3}, \\
\int_{\theta=0}^{2\pi} \int_{r=20a}^{50a} \vert K(\theta , r)\vert r\,dr\,d\theta = 2.7\times 10^{-2} \,.
\end{eqnarray}
These integrals of the Gaussian curvature are at least one order of magnitude smaller than the integrals of the absolute value of the Gaussian curvature in both the core and the rim regions, which again suggests that overall the Gaussian curvature is compensated locally. 

Through Gaussian curvatures, we can confirm the vanishing-mean-curvature phenomenon.  The vanishing of mean curvature $ (C_{rr}+C_{\theta \theta})/2 $ means that the Gaussian curvature $ K $ remains equal to $ C_{\theta \theta}^2 $. Since, as stated in Sec. II, $ C_{\theta \theta}^2 $ at the rim approaches the constant $ K_0 $ as the thickness $ h $ of the sheet goes to zero, the vanishing of the mean curvature would further imply that at the limit $ h=0 $,  $ K = K_0 $ at the rim. In Figure \ref{Kvsh}, there is a clear trend that $ K/K_{0} $ approaches $ 1 $ as $ h $ goes to zero.
  
\section{Discussion}
If an external normal point force is applied to an elastic membrane, positive Gaussian curvature is necessarily created at the forcing point. In general this must be compensated by a region of negative Gaussian curvature elsewhere in the membrane, owing to the Gauss-Bonnet theorem. For the case of the d-cone lattice considered here the content of the theorem is quite precise:  the angular deficits defined above must sum to zero. Thus any deficit due to an external force must be exactly compensated elsewhere in the sheet. This compensation may occur close to the applied force or far from it, depending on the case. The size of the compensating region can depend on the magnitude of the applied force, the forces at distant boundaries and the elastic thickness of the membrane. Similarly the degree of compensation falls off with distance from the source in a way that depends on the case. The nature of compensation for weak point forces is tractable using the linearized von K\'{a}rm\'{a}n equations.

Our interest here is the case of the d-cone, where puzzling nonlinear and nonlocal effects like the vanishing of mean curvature at the supporting rim are seen. As explained above, this vanishing appears to require nonlocal interaction between the rim region and the rest of the system. Any nonlocal compensation of Gaussian curvature would provide a potential mechanism for this nonlocal interaction. However, our numerical analysis shows that the rim gaussian curvature is almost entirely compensated close to the rim. Indeed, the compensation appears to become more local as the elastic thickness of the membrane is reduced. Similar locality appears in the compensation for the central point force. Thus we have found nothing to suggest a coupling of the central region to the rim via the Gauss-Bonnet theorem.

Our findings here are modest and limited. We have only explored one structure: the d-cone, although the locality of compensation is of general interest and relevance. Clearly many variants of the d-cone could be studied, as well as many other cases of point forces on curved and strained membranes. We expect that the increased locality we observed with decreasing thickness obeys some form of scaling behavior, but we have not attempted to determine it numerically or analytically. In order to demonstrate such scaling, any numerical measurement would need to have a finite element scheme with much more dynamic range than the one used here.

The puzzle of vanishing mean curvature at the rim of a d-cone is as open as ever after this study. Our work in preparation investigates the generality of this curvature cancellation and explores simplifying assumptions to enable understanding.

\textbf{Acknowledgments}

The authors would like to thank Konstantin Turitsyn and Enrique Cerda for enlightening discussions. This work was supported in part by the National Science Foundation's MRSEC Program under Grant Number DMR 0820054.

\vspace*{-1ex} \bibliographystyle{apsrev}

\end{document}